\documentstyle[12pt]{article}
\parskip =\medskipamount
\def\be{\begin{equation}}
\def\ee{\end{equation}}
\def\disp{\displaystyle}
\newtheorem{definition}{Definition}

\newcounter{fig}

\begin{document}

\begin{titlepage}
\begin{center}
{\Large\bf Statistics of Reduced Words in Locally Free and
Braid Groups: Abstract Studies and Applications to Ballistic Growth
Model

}
\vspace{.4in}

{\large J. Desbois$^{\dag}$ ~and ~S. Nechaev$^{\ddag\dag}$}
\bigskip

{\sl $^{\dag}$ Institut de Physique Nucl\'eaire, Division de Physique
Th\'eorique},$^*$ \\ {\sl 91406 Orsay Cedex, France}
\medskip

{\sl $^{\ddag}$ L D Landau Institute for Theoretical Physics, \\ 117940,
Moscow, Russia}

\end{center}
\vspace{.5in}

\begin{abstract}
We study numerically and analytically the average length of reduced
(primitive) words in the so-called locally free (${\cal LF}_n(d)$) and
braid ($B_n$) groups. We consider the situations when the letters in the
initial words are drawn either without or with correlations. In latter
case, we show that the average length of the reduced  word can
be increased or lowered depending on the type of correlation. The ideas
developed are used for analytical computation of the average number of
peaks of the surface appearing in some specific ballistic growth model.

\end{abstract}
\vspace{0.2in}

\vspace{0.2in}

\noindent {\sl Submitted to J. Phys. (A): Math. \& Gen.}
\vspace{0.2in}

\vspace{0.5in}

\hrule \footnotesize
$^{*}$ Unit\'e de Recherche des Universit\'es Paris XI et Paris VI
associ\'ee au C.N.R.S.

\end{titlepage}

\tableofcontents

\section*{Introduction}
The paper is devoted to the elaboration of a common new method of analysis
of stable probability distributions in statistical systems of completely
different physical nature, such as: vortices in superconductors, entangled
polymer bunches and open surfaces of growing media. Our paper pursues two
main goals:

(i) To construct a mathematical apparatus to adequately describe the
topological properties of the above physical systems;

(ii) To apply evolved methods to the detailed analysis of specific
physical phenomena.

The development of mathematical methods implies the construction of the
statistical theory of random walks on nonabelian groups
(Refs.\cite{kes}--\cite{den});
while the application of elaborated methods in physics is aimed to answer
the following question (\cite{dena}): how does the change in the topological
state of the system affects its physical properties?

Although the general concepts of the noncommutative probability theory have
been well elaborated in the field-theoretic context, their application in
the related areas of mathematics and physics, such as, for instance,
statistical physics of chain-like objects is highly limited. This state of
affairs can be accounted for by two facts: (a) there is a communication
problem, i.e. the languages used by specialists in topological field theory
and probability theory are completely different at first glance; (b)
physical systems give no evidence how these ideas are reflected in simple
geometrical examples.

So, the present paper is mainly concerned with the probabilistic methods
which allow us to solve the basic problems dealing with the limit
distributions of random walks on some simplest noncommutative groups. To
be more specific, our main goal of the work is as follows: we consider
analytically and numerically the limit behavior of the Markov chains where
the states are randomly taken from some noncommutative finite discrete
group. In particular, we restrict ourselves with the so-called {\it braid}
($B_{n+1}$)  and {\it locally free} (${\cal LF}_{n+1}$) groups---see the
definitions in the section \ref{subsec:1.2}. The preliminary results
concerning the words statistics in the locally free groups appeared in
recent works \cite{nevegr,den}.

The reason of our investigations is forced by the following real physical
motivations:

(a) The nematic-type ordering in bunches of entangled polymers as well as
the consideration of thermodynamic properties of uncrossible vortex lines
immediately turn us to studying of a statistics of chain-like objects with
nonabelian topology. The main reason of investigation concerns the
construction of the basis of the mean--field--like theory of fluctuating
entangled chains in 1+1 dimensions on the basis of our knowledges about
statistics of Markov chains on braid and locally free groups. A forthcoming
publication will be devoted to examination of the influence of
topological constraints in the standard nematic--like phase transition in
the bunch of "braided polymers" \cite{dena}.

(b) It has been realized that the ballistic--type growth of some amount of
deposit in a box and the investigation of the shape of the  surface can be
easily translated into the language of random walks over the elements of
some noncommutative group. The corresponding model is considered in the
section \ref{sec:4} of the present work.

\section{Basic Definitions}\label{sec:1}
We begin with the investigation of the probabilistic properties of Markov
chains on simplest noncommutative groups. In the most general way the
problem can be formulated as follows (see also \cite{nevegr}).

Take a discrete group ${\cal G}_{n+1}$ constructed by the finite number
of generators $\{g_1,\ldots, \break g_n\}$. Any arbitrary sequence of
generators we call the {\it initial word}. The {\it length}, $N$, of this
word is the total number of used generators ("letters"), whereas the {\it
length} of the {\it reduced} (or {\it primitive}) {\it word}, $\mu$, is
the shortest noncontractible length of the word which remains after
applying of all possible group relations.

Later on we mainly use the rescaled variables $N'\equiv N/n$ and
$\mu'\equiv \mu/n$ instead of $N$ and $\mu$ and consider the situation
$n\gg 1$ neglecting the "edge effects".

The most attention in the following is paid to the computation of the mean
length, $\left<\mu'(N')\right>$, averaged over various distributions of
initial words belonging to the group ${\cal G}_{n+1}$ (${\cal G}_{n+1}$
is either "locally free" or braid group).

\subsection{Random Walks Over Group Elements}\label{subsec:1.1}
Take the group ${\cal G}_{n+1}$. Let $p$ be some distribution on the set
$\{g_1,\ldots, g_n, g_1^{-1}, \ldots,g_n^{-1}\}$. For convenience
we call $h_j\equiv g_i$ for $j=i$ and $h_j\equiv g_i^{-1}$ for $j\equiv
i+n$. We construct the (right-hand) random walk (the random word) on ${\cal
G}_{n+1}$ with a transition measure, $p$, i.e. we add with the
probability $p$ the element $h_{\alpha_{N+1}}$ to the given word $W_N=
h_{\alpha_1} h_{\alpha_2}\ldots h_{\alpha_N}$ from the right-hand
side\footnote{Analogously we can construct the left-hand side random walk
on the group ${\cal G}_{n+1}$.}.

The random word $W$ formed by $N$ letters taken with
the probability distribution $p$ from the set $\{g_1,\ldots, g_n,
g_1^{-1},\ldots, g_n^{-1}\}$ is called the {\it initial word}
of the length $N$ on the group ${\cal G}_{n+1}$.

We distinguish below between the following three situations:
\medskip

\noindent {\sc 1. Drawing words without any correlations
("standard case").}

The probability distribution $p$ is uniform, i.e. $p=\frac{1}{2n}$ on the
set $\{g_1,\ldots,$ $g_n, g_1^{-1},\ldots,\break g_n^{-1}\}$.

Besides this standard case, we consider also two extreme situations of
words construction, hereafter refered as "weak (strong) correlations".
\medskip

\noindent {\sc 2. Drawing words with weak correlations (regime
"A").}

Suppose that in the initial word, $W_N$, we have for the last letter
$h_{\alpha_N}=g_k$ or $g_k^{-1}$. Then we add the next $(N+1)$th letter
$h_{\alpha_{N+1}}$ with the following probabilities:
\be
h_{\alpha_{N+1}}=\left\{\begin{array}{cl}
\disp g_k^{\pm 1} & \mbox{with the probability $q_A$} \\
\mbox{\it any other letter} & \mbox{with the probability $p_A$}
\end{array}\right.
\ee
The normalisation reads:
\be\label{nor:1}
2q_A+2(n-1)p_A=1
\ee
\medskip

\noindent {\sc 3. Drawing words with strong correlations (regime
"B").}

Suppose again that in the initial word, $W_N$, we have for the last letter
$h_{\alpha_N}=g_k$ or $g_k^{-1}$. Then we add the next $(N+1)$th letter
$h_{\alpha_{N+1}}$ with the following probabilities:
\be
h_{\alpha_{N+1}}=\left\{\begin{array}{cl}
\disp g_{k\pm 1}^{\pm 1} & \mbox{with the probability $q_B$} \\
\mbox{\it any other letter} & \mbox{with the probability $p_B$}
\end{array}\right.
\ee
The normalisation in this case reads:
\be\label{nor:2}
4q_B+2(n-2)p_B=1
\ee

In particular, we show below that in the case "A" the length of the
reduced (primitive) word decreases when $q_A$ is increased, while in
the case "B" the length of the reduced word increases when $q_B$ is
increased.

The absence of correlations, in both cases "A" and "B", means setting
$q_{A,B}=p_{A,B}=\frac{1}{2n}$. Thus, in the limit $n\gg 1$, the
"standard case" is recovered, formally, by setting $q_{A,B} = 0$ in
the equations.

The investigation of such correlations is necessary in view of future
physical applications \cite{dena}, especially when we are dealing with the
polymers entanglements. Indeed, if we think of ${\cal G}_{n+1}$ as of a braid
group, the weak ("A") and strong ("B") correlation regimes will correspond,
respectively, to the weak and strong "entanglement regimes".

\subsection{Braid and "Locally Free" Groups}\label{subsec:1.2}
We are aimed to study the asymptotics of the limit distributions of Markov
chains on the braid group $B_{n+1}$. For the case $n=2$ the problem has
been solved in \cite{nevegr}, where the limit probability distribution as
well as the conditional limit probability distribution of "brownian
bridges" on the group $B_3$ has been derived. For $n>2$ this problem is
unsolved yet. However we can extract some reliable estimations for the limit
behavior of Markov chains on $B_{n+1}$ considering the random walks on
so-called "locally free groups" \cite{vershik,nevegr,den}.

\noindent{\sc Braid Group}. The braid group $B_{n+1}$ of $n+1$ strings has
$n$ generators $\{\sigma_1,\sigma_2,\ldots,\sigma_n\}$ with the following
relations:
\be \label{2:1}
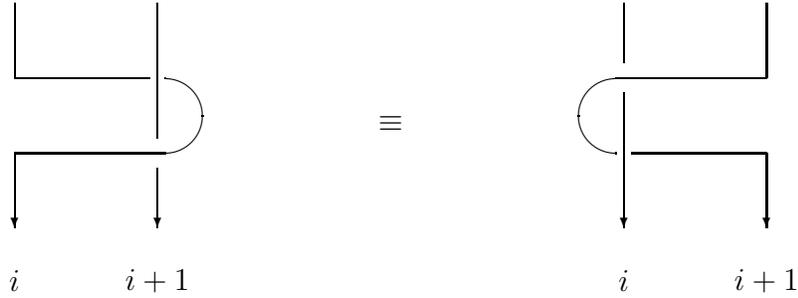
\begin{array}{ll}
\sigma_i\sigma_{i+1}\sigma_i = \sigma_{i+1}\sigma_i\sigma_{i+1}
& \qquad (1\le i < n) \medskip \\
\sigma_i\sigma_j=\sigma_j\sigma_i & \qquad (|i-j|\ge 2) \medskip \\
\sigma_i\sigma_i^{-1}=\sigma_i^{-1}\sigma_i=\hat{e} &
\end{array}
\ee

\noindent Let us mention that:

\noindent -- The word written in terms of "letters"---generators from
the set $\{\sigma_1,\ldots, \sigma_n,\sigma_1^{-1},\ldots,\break
\sigma_n^{-1}\}$ gives a particular {\it braid}. Schematically, the
generators $\sigma_i$ and $\sigma_i^{-1}$ could be represented as follows:

\bigskip

\unitlength=1.00mm
\special{em:linewidth 0.5pt}
\linethickness{0.5pt}
\hspace{2.5cm}
\begin{picture}(100.00,50.00)
\put(5.00,50.00){\line(0,-1){15.00}}
\put(30.00,50.00){\line(1,-1){5.00}}
\put(40.00,40.00){\line(1,-1){5.00}}
\put(45.00,50.00){\line(-1,-1){15.00}}
\put(-10.00,50.00){\line(0,-1){15.00}}
\put(70.00,50.00){\line(0,-1){15.00}}
\put(85.00,50.00){\line(0,-1){15.00}}
\put(5.00,17.00){\line(0,-1){15.00}}
\put(-10.00,17.00){\line(0,-1){15.00}}
\put(70.00,17.00){\line(0,-1){15.00}}
\put(85.00,17.00){\line(0,-1){15.00}}
\put(30.00,17.00){\line(1,-1){15.00}}
\put(45.00,17.00){\line(-1,-1){5.00}}
\put(35.00,7.00){\line(-1,-1){5.00}}
\put(-10.00,28.00){\makebox(0,0)[cc]{$1$}}
\put(5.00,28.00){\makebox(0,0)[cc]{$2$}}
\put(17.00,35.00){\makebox(0,0)[cc]{$\ldots$}}
\put(17.00,28.00){\makebox(0,0)[cc]{$\ldots$}}
\put(30.00,28.00){\makebox(0,0)[cc]{$i$}}
\put(45.00,28.00){\makebox(0,0)[cc]{$i+1$}}
\put(57.00,35.00){\makebox(0,0)[cc]{$\ldots$}}
\put(57.00,28.00){\makebox(0,0)[cc]{$\ldots$}}
\put(70.00,28.00){\makebox(0,0)[cc]{$n$}}
\put(85.00,28.00){\makebox(0,0)[cc]{$n+1$}}
\put(100.00,40.00){\makebox(0,0)[ll]{$\large =\sigma_i$}}
\put(-10.00,-5.00){\makebox(0,0)[cc]{$1$}}
\put(5.00,-5.00){\makebox(0,0)[cc]{$2$}}
\put(17.00,2.00){\makebox(0,0)[cc]{$\ldots$}}
\put(17.00,-5.00){\makebox(0,0)[cc]{$\ldots$}}
\put(30.00,-5.00){\makebox(0,0)[cc]{$i$}}
\put(45.00,-5.00){\makebox(0,0)[cc]{$i+1$}}
\put(57.00,2.00){\makebox(0,0)[cc]{$\ldots$}}
\put(57.00,-5.00){\makebox(0,0)[cc]{$\ldots$}}
\put(70.00,-5.00){\makebox(0,0)[cc]{$n$}}
\put(85.00,-5.00){\makebox(0,0)[cc]{$n+1$}}
\put(100.00,7.00){\makebox(0,0)[ll]{$\large =\sigma_i^{-1}$}}
\end{picture}

\vspace{0.3in}

\noindent -- The {\it length}, $N$, of the braid is the total number of
used letters while the {\it minimal irreducible length}, $\mu$, is the
shortest noncontractible length of a particular braid remaining after all
possible group relations Eq.(\ref{2:1}) are applied. Diagramatically, the
braid can be represented as a set of crossed strings going from the top to
the bottom after "gluing" the braid generators.

\noindent -- The closed braid appears after gluing the "upper" and the
"lower" free ends of the braid on the cylinder.

\noindent -- Any braid corresponds to some knot or link. So, there is a
principal possibility to use the braid group representation for the
construction of topological invariants of knots and links, but the
correspondence of braids and knots is not mutually single valued and each
knot or link can be represented by an infinite series of different braids.
\bigskip

\noindent {\sc Locally Free Group}. The group ${\cal LF}_{n+1}(d)$ is called
{\it locally free} if the generators, $\{\sigma_1,\ldots,\sigma_n\}$ obey the
following commutation relations:
\begin{itemize}
\item[(a)] Each pair $(\sigma_j, \sigma_k)$ generates the free subgroup of the group
${\cal LF}_{n+1}$ if $|j-k|<d$;
\item[(b)] $\sigma_j \sigma_k=\sigma_k \sigma_j$ for  $|j-k|\ge d$
\end{itemize}
We will be concerned mostly with the case $d=2$ for which we define
${\cal LF}_{n+1}(2)\equiv {\cal LF}_{n+1}$.

\noindent The graphical representation of generators $\sigma_i$ and
$\sigma_i^{-1}$ is rather similar to that of braid group:

\bigskip
\unitlength=1.00mm
\special{em:linewidth 0.5pt}
\linethickness{0.5pt}
\hspace{2.5cm}
\begin{picture}(100.00,50.00)
\put(5.00,50.00){\line(0,-1){15.00}}
\put(-10.00,50.00){\line(0,-1){15.00}}
\put(70.00,50.00){\line(0,-1){15.00}}
\put(85.00,50.00){\line(0,-1){15.00}}
\put(5.00,17.00){\line(0,-1){15.00}}
\put(-10.00,17.00){\line(0,-1){15.00}}
\put(70.00,17.00){\line(0,-1){15.00}}
\put(85.00,17.00){\line(0,-1){15.00}}
\put(-10.00,28.00){\makebox(0,0)[cc]{$1$}}
\put(5.00,28.00){\makebox(0,0)[cc]{$2$}}
\put(17.00,35.00){\makebox(0,0)[cc]{$\ldots$}}
\put(17.00,28.00){\makebox(0,0)[cc]{$\ldots$}}
\put(30.00,28.00){\makebox(0,0)[cc]{$i$}}
\put(45.00,28.00){\makebox(0,0)[cc]{$i+1$}}
\put(57.00,35.00){\makebox(0,0)[cc]{$\ldots$}}
\put(57.00,28.00){\makebox(0,0)[cc]{$\ldots$}}
\put(70.00,28.00){\makebox(0,0)[cc]{$n$}}
\put(85.00,28.00){\makebox(0,0)[cc]{$n+1$}}
\put(100.00,40.00){\makebox(0,0)[ll]{$\large =\sigma_i$}}
\put(-10.00,-5.00){\makebox(0,0)[cc]{$1$}}
\put(5.00,-5.00){\makebox(0,0)[cc]{$2$}}
\put(17.00,2.00){\makebox(0,0)[cc]{$\ldots$}}
\put(17.00,-5.00){\makebox(0,0)[cc]{$\ldots$}}
\put(30.00,-5.00){\makebox(0,0)[cc]{$i$}}
\put(45.00,-5.00){\makebox(0,0)[cc]{$i+1$}}
\put(57.00,2.00){\makebox(0,0)[cc]{$\ldots$}}
\put(57.00,-5.00){\makebox(0,0)[cc]{$\ldots$}}
\put(70.00,-5.00){\makebox(0,0)[cc]{$n$}}
\put(85.00,-5.00){\makebox(0,0)[cc]{$n+1$}}
\put(100.00,7.00){\makebox(0,0)[ll]{$\large =\sigma_i^{-1}$}}
\put(30.00,38.00){\line(1,0){13.00}}
\put(30.00,50.00){\line(0,-1){5.00}}
\put(30.00,38.00){\line(0,-1){3.00}}
\put(45.00,50.00){\line(0,-1){3.00}}
\put(45.00,42.00){\line(0,-1){7.00}}
\put(30.00,13.00){\line(1,0){13.00}}
\put(30.00,17.00){\line(0,-1){4.00}}
\put(30.00,6.00){\line(0,-1){4.00}}
\put(45.00,17.00){\line(0,-1){9.00}}
\put(45.00,5.00){\line(0,-1){3.00}}
\put(47.00,41.48){\oval(5.06,6.97)[r]}
\put(47.00,45.00){\line(-1,0){16.95}}
\put(47.00,9.45){\oval(5.06,6.97)[r]}
\put(47.00,6.00){\line(-1,0){16.90}}
\end{picture}

\vspace{0.3in}

It is easy to understand that the following geometrical identity is valid:

\bigskip
\unitlength=1.00mm
\special{em:linewidth 0.4pt}
\linethickness{0.4pt}
\hspace{1cm}
\begin{picture}(110.00,40.00)
\put(110.00,40.00){\line(0,-1){10.00}}
\put(110.00,20.00){\vector(0,-1){10.00}}
\put(10.00,40.00){\line(0,-1){10.00}}
\put(10.00,20.00){\vector(0,-1){10.00}}
\put(29.00,40.00){\line(0,-1){18.00}}
\put(29.00,18.00){\vector(0,-1){8.00}}
\put(91.00,40.00){\line(0,-1){8.00}}
\put(91.00,28.00){\vector(0,-1){18.00}}
\put(60.00,24.00){\makebox(0,0)[cc]{$\Large \equiv$}}
\put(90.00,25.00){\oval(10.00,10.00)[l]}
\put(110.00,30.00){\line(-1,0){20.00}}
\put(30.00,25.00){\oval(10.00,10.00)[r]}
\put(92.00,20.00){\line(1,0){18.00}}
\put(30.00,20.00){\line(-1,0){20.00}}
\put(10.00,30.00){\line(1,0){18.00}}
\put(10.00,3.00){\makebox(0,0)[cc]{$i$}}
\put(29.00,3.00){\makebox(0,0)[cc]{$i+1$}}
\put(91.00,3.00){\makebox(0,0)[cc]{$i$}}
\put(110.00,3.00){\makebox(0,0)[cc]{$i+1$}}
\end{picture}

\noindent hence, it is unnecessary to distinguish between "left" and
"right" operators $\sigma_i$.

It can be seen that the only difference between the braid and locally
free groups consists in elimination of the Yang-Baxter relations (first line
in Eq.(\ref{2:1})).

\section{Random Walks without Correlations on Locally Free and Braid
groups}\label{sec:2}
It has been shown in papers \cite{kes},\cite{nesin}--\cite{nesem} that for
the free group (i.e. for the group without any commutation relations
among generators) the problem of the limit distribution of Markov chains
can be mapped to the investigation of statistics of random walks on a
simply connected tree. In the case of locally free groups or braid groups
the more complicated  structure does not allow us to use this simple
geometrical image directly.

\subsection{The Locally Free Group ${\cal LF}_{n+1}(2)$}
\label{subsec:2.1}

Let us begin with the following example:

\noindent {\bf Example 1.}

Suppose that the $N$-letter initial word leads to the following
reduced word:
$$
\sigma_1^{-1}\sigma_2\sigma_1\sigma_4\sigma_2^{-1}\sigma_7\sigma_3
\sigma_5^{-1}\sigma_3\sigma_8^{-1}
$$
Now, if we add randomly a new letter from the right hand side, it is easy
to see that only $\sigma_3$, $\sigma_5^{-1}$ or $\sigma_8^{-1}$ can be
reduced (for instance, $\sigma_7$ cannot be reduced even if, by chance, we
add $\sigma_7^{-1}$ because this generator cannot pass through
$\sigma_8^{-1},\ldots$).

\begin{definition}
The set of letters which we can reduce in the given primitive word by
adding one extra letter from the right--hand side we call {\bf the
set of reducible letters}, $I$.
\end{definition}
The number of letters belonging to $I$ we denote as $\eta$.

In the above example $I=\{\sigma_3,\sigma_5^{-1}, \sigma_8^{-1}\}$
and $\eta=3$. Generally speaking, $\eta'\equiv \eta /n$ is a random
variable, the probability distribution of which {\it a priori} depends both
on $N'\equiv N/n$ and $n$.

It is noteworthy to mention the following basic properties of the set
$I$:
\begin{itemize}
\item[{\rm(i)}] If $\sigma_i^{\pm 1}$ belongs to $I$ then
$\sigma_i^{\mp 1}$ does not belong to $I$;
\item[{\rm(ii)}] If $\sigma_i^{\pm 1}$ belongs to $I$ then
$\sigma_{i+1}^{\pm 1}$, $\sigma_{i+1}^{\mp 1}$, $\sigma_{i-1}^{\pm 1}$ and
$\sigma_{i-1}^{\mp 1}$ do not belong to $I$, i.e. {\bf all the elements of
$I$ must commute}.
\end{itemize}
On the basis of (i) and (ii) we can easily deduce that $0\le\eta\le n/2$
($\eta=0$ corresponds to a completely reduced word, i.e. $\mu=0$).
\medskip

The set $I$ allows the following very useful geometrical
interpretation. Take $n$ boxes (labelled as $k=1,\ldots,n$) as displayed
below:

\vspace{0.3in}
\unitlength=1mm
\special{em:linewidth 0.5pt}
\linethickness{0.5pt}
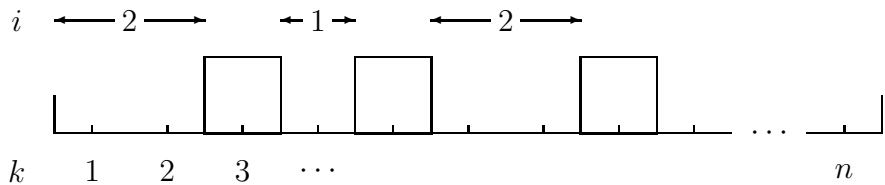
\hspace{0.5cm}
\begin{picture}(125.00,25.00)
\put(20.00,10.00){\line(0,1){1.00}}
\put(30.00,10.00){\line(0,1){1.00}}
\put(40.00,10.00){\line(0,1){1.00}}
\put(50.00,10.00){\line(0,1){1.00}}
\put(60.00,10.00){\line(0,1){1.00}}
\put(70.00,10.00){\line(0,1){1.00}}
\put(80.00,10.00){\line(0,1){1.00}}
\put(90.00,10.00){\line(0,1){1.00}}
\put(100.00,10.00){\line(0,1){1.00}}
\put(15.00,15.00){\line(0,-1){5.00}}
\put(105.00,10.00){\line(0,0){0.00}}
\put(55.00,10.00){\framebox(10.00,10.00)[cc]{}}
\put(85.00,10.00){\framebox(10.00,10.00)[cc]{}}
\put(15.00,10.00){\line(1,0){90.00}}
\put(110.00,10.00){\makebox(0,0)[cc]{$\cdots$}}
\put(115.00,10.00){\line(1,0){10.00}}
\put(125.00,10.00){\line(0,0){0.00}}
\put(125.00,10.00){\line(0,1){5.00}}
\put(35.00,10.00){\framebox(10.00,10.00)[cc]{}}
\put(20.00,5.00){\makebox(0,0)[cc]{1}}
\put(30.00,5.00){\makebox(0,0)[cc]{2}}
\put(40.00,5.00){\makebox(0,0)[cc]{3}}
\put(50.00,5.00){\makebox(0,0)[cc]{$\cdots$}}
\put(120.00,10.00){\line(0,1){1.00}}
\put(120.00,5.00){\makebox(0,0)[cc]{$n$}}
\put(23.00,25.00){\vector(-1,0){8.00}}
\put(27.00,25.00){\vector(1,0){8.00}}
\put(48.00,25.00){\vector(-1,0){3.00}}
\put(52.00,25.00){\vector(1,0){3.00}}
\put(73.00,25.00){\vector(-1,0){8.00}}
\put(77.00,25.00){\vector(1,0){8.00}}
\put(25.00,25.00){\makebox(0,0)[cc]{2}}
\put(50.00,25.00){\makebox(0,0)[cc]{1}}
\put(75.00,25.00){\makebox(0,0)[cc]{2}}
\put(10.00,25.00){\makebox(0,0)[cc]{$i$}}
\put(10.00,5.00){\makebox(0,0)[cc]{$k$}}
\end{picture}

\vspace{-0.3cm}
\noindent The box $k$ is empty except if $\sigma_k^{\pm 1}$ belongs to
$I$. In the given example only boxes $3$, $5$, $8$ are occupied. From
the properties (i)--(ii) of the set $I$ we deduce that two neighboring
boxes cannot be occupied.

Generally, $I$ is described by occupied boxes separated by a sequence
of $i$ ($i\ge 1$) empty boxes. Let $n_i$ being the number of such sequences
of length $i$. Neglecting the edge effects (i.e. for $n\gg 1$), we get
the following rules:
\be \label{dd:1}
\sum_{i\ge 1}n_i=\eta
\ee and \be \label{dd:2}
\sum_{i\ge 1}i\;n_i=n-\eta
\ee

Consider now the evolutions of the reduced word (length $\mu$) and
of the set $I$ (length $\eta$) when we add randomly a letter from the
right hand side: $N\to N+1$ (i.e. $N'\to N'+1/n$). (Apparently the evolutions
of the reduced word and of the set $I$ are correlated). Two possibilities
 can occur:
$$
\left\{\begin{array}{ll}
\disp \Delta\mu'=+\frac{1}{n} & \mbox{for the "increase" process
($\mu\to\mu+1$)} \medskip \\
\disp \Delta\mu'=-\frac{1}{n} & \mbox{for the "decrease" process
($\mu\to\mu-1$)}
\end{array}\right.
$$
where $\Delta\mu'$ stands for the increment $\mu'(N'+1/n)-\mu'(N')$.

We consider the "increase" and "decrease" processes separately.

\noindent {\bf 1. The "increase" process.}

\bigskip

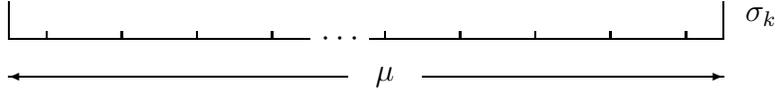
\unitlength=1.00mm
\special{em:linewidth 0.5pt}
\linethickness{0.5pt}
\begin{picture}(150.00,20.00)
\hspace{-1cm}
\put(35.00,15.00){\line(0,-1){5.00}}
\put(40.00,10.00){\line(0,1){1.00}}
\put(50.00,10.00){\line(0,1){1.00}}
\put(60.00,10.00){\line(0,1){1.00}}
\put(70.00,10.00){\line(0,1){1.00}}
\put(79.00,10.00){\makebox(0,0)[cc]{$\cdots$}}
\put(85.00,10.00){\line(0,1){1.00}}
\put(95.00,10.00){\line(0,1){1.00}}
\put(105.00,10.00){\line(0,1){1.00}}
\put(115.00,10.00){\line(0,1){1.00}}
\put(125.00,10.00){\line(0,1){1.00}}
\put(130.00,15.00){\line(0,-1){5.00}}
\put(135.00,13.00){\makebox(0,0)[cc]{$\sigma_k$}}
\put(85.00,5.00){\makebox(0,0)[cc]{$\mu$}}
\put(35.00,10.00){\line(1,0){40.00}}
\put(83.00,10.00){\line(1,0){47.00}}
\put(80.00,5.00){\vector(-1,0){45.00}}
\put(90.00,5.00){\vector(1,0){40.00}}
\end{picture}

It is easy to see that the added letter will necessarily belong  to the new
set $I$.  However, it does
not mean at all that, in this case, we {\it automatically} have $\eta'\to
\eta'+1/n$. Actually, $\eta'$ can stay unchanged or can be changed by $\pm
1/n$.

The latter point becomes more clear if we come back to Example 1 where
$I=\{\sigma_3,\sigma_5^{-1}, \sigma_8^{-1}\}$. We have the following
choice:
\begin{itemize}
\item If we add $\sigma_3$ (or $\sigma_5^{-1}$ or $\sigma_8^{-1}$),
 then the set $I$ (and, hence, $\eta'$) remains to be unchanged.
\item If we add $\sigma_6$, then $\eta'$ is still unchanged: $I$
becomes the new set $\{\sigma_3,\sigma_6,\sigma_8^{-1}\}$, the letter
$\sigma_6$ has replaced $\sigma_5^{-1}$ in the set $I$. The same
occurs for $\eta'$ if, instead of $\sigma_6$, we add $\sigma_6^{-1}$ or
$\sigma_7^{\pm 1}$, and so on...
\item If we add $\sigma_4$, then $I$ becomes the new set $\{\sigma_4,
\sigma_8^{-1}\}$ ($\sigma_4$ erases $\sigma_3$ and $\sigma_5^{-1}$
from $I$) and, consequently, $\eta'\to\eta'-1/n$ (same change occurs
for $\eta'$ if we add $\sigma_4^{-1}$).
\item If we add $\sigma_{10}$, then $I$ becomes $\{\sigma_3,
\sigma_5^{-1}, \sigma_8^{-1},\sigma_{10}\}$ and $\eta'\to\eta'+1/n$ (the
same is happened if we add $\sigma_1^{\pm 1}$ or $\sigma_{10}^{-1}$,
and so on...).
\end{itemize}

These considerations can be generalized and careful inspection leads to the
following rules for the increasing process ($\Delta N'=+1/n$,
$\Delta\mu'=+1/n$):
\be \label{dd:30}
\left\{\begin{array}{ll}
\Delta\eta'=0 & \mbox{occurs with probability $\disp \Pi_0=
\frac{1}{2n}\left(\eta+4\sum_{i\ge 2}n_i\right)$} \medskip \\
\disp \Delta\eta'=-\frac{1}{n} & \mbox{occurs with probability
$\disp \Pi_{-}=\frac{1}{2n}\,2n_1$} \medskip \\
\disp \Delta\eta'=+\frac{1}{n} & \mbox{occurs with probability
$\disp \Pi_{+}=\frac{1}{2n}\,2\sum_{i\ge 3}n_i(i-2)$}
\end{array}\right.
\ee

With the help of Eqs.(\ref{dd:30}) we get as expected, for the
total probability, $\Pi_1$, of the increasing process:
\be \label{dd:4}
\Pi_1=\Pi_0 + \Pi_{-} + \Pi_{+} = 1-\frac{\eta'}{2}
\ee
 From the inequality $0\le\eta'\le 1/2$ derived above immediately follows
 that $3/4\le\Pi_1\le 1$. For the corresponding average change of $\eta'$ we
 have:
$$
\frac{\left<\Delta_1\eta'\right>}{\Delta N'}=-\frac{n_1}{n}+
\frac{1}{n}\sum_{i\ge 3}n_i(i-2)=\frac{1}{n}\sum_{i\ge 1}n_i(i-2)=
\frac{n-3\eta}{n}
$$
so, we arrive at the following equation
\be \label{dd:5}
\left<\Delta_1\eta'\right>=(1-3\left<\eta'\right>)\Delta N'
\ee
where $\left<\ldots\right>$ stands for averaging over the set of all
initial words with $N'$ fixed.
\medskip

\noindent {\bf 2. The "decrease" process.}

Now we compute the change $\left<\Delta_2\eta'\right>$ for the reducing
process, i.e. when $N'\to N'+1/n$ and $\mu'\to\mu'-1/n$. It occurs with the
probability
$$
\Pi_2 \equiv 1-\Pi_1=\frac{\eta'}{2}
$$
In this operation, a letter of the set $I$ is erased and, again, we have
$\Delta\eta'=0$ or $\pm 1/n$. Recall that all the elements of $I$ commute.
So, the erased letter (here $\sigma_k$) can always been considered as the
last one:

\bigskip

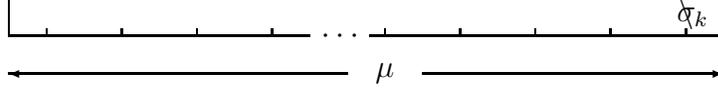
\unitlength=1.00mm
\special{em:linewidth 0.5pt}
\linethickness{0.5pt}
\begin{picture}(150.00,20.00)
\hspace{-1cm}
\put(35.00,15.00){\line(0,-1){5.00}}
\put(40.00,10.00){\line(0,1){1.00}}
\put(50.00,10.00){\line(0,1){1.00}}
\put(60.00,10.00){\line(0,1){1.00}}
\put(70.00,10.00){\line(0,1){1.00}}
\put(79.00,10.00){\makebox(0,0)[cc]{$\cdots$}}
\put(85.00,10.00){\line(0,1){1.00}}
\put(95.00,10.00){\line(0,1){1.00}}
\put(105.00,10.00){\line(0,1){1.00}}
\put(115.00,10.00){\line(0,1){1.00}}
\put(125.00,10.00){\line(0,1){1.00}}
\put(130.00,15.00){\line(0,-1){5.00}}
\put(125.00,13.00){\makebox(0,0)[cc]{$\sigma_k\hspace{-0.4cm}\backslash$}}
\put(85.00,5.00){\makebox(0,0)[cc]{$\mu$}}
\put(35.00,10.00){\line(1,0){40.00}}
\put(83.00,10.00){\line(1,0){47.00}}
\put(80.00,5.00){\vector(-1,0){45.00}}
\put(90.00,5.00){\vector(1,0){40.00}}
\end{picture}

From this point of view, the decrease process $(N\to N+1$ , $\mu\to\mu -1)$
is rigorously the inverse of the increase one $(N-1\to N$ , $\mu -1\to\mu )$.
Thus, weighting each  process with its actual probability, we get the
equation:
\be \label{dd:6}
\left<\Delta_2\eta'\right>=-\left<\Delta_1\eta'\, \frac{\Pi_2}{\Pi_1}
\right>
\ee
where corrections of order $1/n$ are neglected.

Collecting the "increase" and "decrease" processes together, we obtain:
$$
\frac{\left<\Delta\eta'\right>}{\Delta N'}=
\left<(1-3\eta')\left(1-\frac{\Pi_2}{\Pi_1}\right)\right>
$$
and
$$
\frac{\left<\Delta\mu'\right>}{\Delta N'}=\Pi_1-\Pi_2
$$

We arrive in the limit $N\gg 1;\; \mu\gg 1$ at the following differential
equations:
\be \label{dd:7}
\frac{d\left<\eta'\right>}{dN'}=\left<(1-3\eta')
\frac{(1-\eta')}{(1-\eta'/2)}\right>
\ee
and
\be \label{dd:8}
\frac{d\left<\mu'\right>}{dN'}=1-\left<\eta'\right>
\ee
It should be stressed that Eq.(\ref{dd:8}) together with the inequality
$\eta'\le 1/2$ imply that $\disp \frac{\left<\mu\right>}{N}\ge 1/2$.

We can get rid of the brackets in
Eqs.(\ref{dd:7})--(\ref{dd:8}) when $n\gg 1$. To show that, let us compute
the probability distribution of $\eta'$, $P(N',\eta')$. The function
$P(N',\eta')$ satisfies the following recursion relation
\be \label{dd:9}
\left\{\begin{array}{l}
P(N'+1/n,\eta')=P_0\,P(N',\eta')+P_1\,P(N',\eta'-1/n)+
P_2\,P(N',\eta'+1/n) \medskip \\
P(0,\eta')=\delta(\eta')
\end{array}\right.
\ee
where
$$
\begin{array}{c}
\disp P_0=\Pi_0\left(1+\frac{\Pi_2}{\Pi_1}\right); \qquad
\disp P_1=\Pi_{+}+\Pi_{-}\frac{\Pi_2}{\Pi_1}; \qquad
\disp P_2=\Pi_{-}+\Pi_{+}\frac{\Pi_2}{\Pi_1}
\end{array}
$$
are transition rates. Expanding Eq.(\ref{dd:9}) to the lowest order in
$1/n$ we get
\be \label{dd:10}
\frac{\partial P}{\partial N'}=
(P_2-P_1)\frac{\partial P}{\partial\eta'}+ \frac{1}{2n}(P_1+P_2)
\frac{\partial^2 P}{\partial\eta'^2}+O\left(\frac{1}{n}\right)
\ee
When $n\to\infty$, the diffusion term becomes negligible and the equation
becomes deterministic. Then, the distribution function $P$ acquires zero's
width, hence $\eta'$ is peaked at its average value. The same would be true
for $\mu'$ but not for variables $\eta$ and $\mu$ (for which a
non-vanishing width is expected).

From now on, as far as only $\eta'$ and $\mu'$ are concerned, we
systematically omit the brackets. Solving Eq.(\ref{dd:7}) we get:
\be
\label{dd:11} \frac{1-\eta'}{(1-3\eta')^{5/3}}=e^{4N'}
\ee
Using Eq.(\ref{dd:8}), we get $\mu'$ as a function of $N'$.

The comparison with the numerical simulations is displayed in the upper part
of Fig.\ref{fig:1} (the full curve: Eqs.(\ref{dd:8}), (\ref{dd:11}); the
points: simulations with $n=100$).

We observe at small $N'$ that $\left<\mu\right>\simeq N$, i.e. practically
no reduction occurs because the words are too short compared to the set of
available letters and we have only little chance to draw, in the same word, a
given generator and its inverse. On the other hand, taking the limit $N'\gg
1$ in Eqs.(\ref{dd:8})--(\ref{dd:11}) we arrive at \cite{den} :
\be
\label{dd:12} \eta'=\frac{1}{3} \quad \mbox{and}\quad
\frac{\left<\mu\right>}{N}=\frac{2}{3}
\ee

\subsection{The Locally Free Group ${\cal LF}_{n+1}(d)$ for $d\ge 2$}
\label{subsec:2.2}
The ideas developed above can be extended to the general case---the
group ${\cal LF}_{n+1}(d)$ with $d\ge 2$. It is just worthwile to mention
the following simple fact. The generator $\sigma_k$ erases all $\sigma_j$'s
with $j=k-(d-1),\ldots,k-1,k+1,\ldots,k+(d-1)$ from the set $I$. In
other words, $\sigma_k$ "screens" all the generators in a zone of extension
$2(d-1)$ around itself. This point of view is especially useful when
we treat the correlations ("B").

In the case of the group ${\cal LF}_{n+1}(d)$ Eqs.(\ref{dd:30}) become:
\be
\left\{\begin{array}{ll}
\Delta\eta'=0 & \mbox{occurs with $\disp\; \Pi_0= \frac{\eta'}{2}+
\sum_{i=d-1}^{2(d-1)} \frac{4n_i(i-(d-1))}{2n}+
\sum_{i>2(d-1)}\frac{4n_i(d-1)}{2n}$} \medskip \\
\disp \Delta\eta'=+\frac{1}{n} & \mbox{occurs with $\disp\; \Pi_{+}=
\sum_{i>2(d-1)} \frac{2n_i(i-2(d-1))}{2n}$} \medskip \\ \disp
\Delta\eta'=-\frac{1}{n} & \mbox{occurs with $\disp\; \Pi_{-}=
\sum_{i=d-1}^{2(d-1)} \frac{2n_i(2(d-1)-i)}{2n}$}
\end{array} \right.
\ee
while Eq.(\ref{dd:8}) remains to be unchanged.

The solution of Eq.(\ref{dd:7}) reads now
\be \label{rr:1}
\frac{1-\eta'}{\left(1-(2d-1)\eta'\right)^{\frac{4d-3}{2d-1}}}=
e^{4(d-1)N'}
\ee
(compare to Eq.(\ref{dd:11})). Asymptotically we get:
\be \label{rr:2}
\eta'=\frac{1}{2d-1} \quad \mbox{and} \quad
\frac{\left<\mu\right>}{N}=\frac{2d-2}{2d-1}
\ee
It is easy to check that Eq.(\ref{dd:12}) is recovered for $d=2$ (see
also \cite{den}).

\subsection{The Braid Group $B_{n+1}$} \label{subsec:2.3}
Comparing the groups $B_{n+1}$ and ${\cal LF}_{n+1}(2)$ we could
see that $\eta'$ in Eq.(\ref{dd:8}) has to be replaced by some $\eta''$
($>\eta'$) in order to take into account the additional braiding relations
Eq.(\ref{2:1}).

For each $\sigma_k$ belonging to $I$, we can get additionnal reduction
if and only if we can build at the end of the reduced word a sequence of
letters like $\sigma_k\sigma_{k+1}\sigma_k$. Then the braiding relation
Eq.(\ref{2:1}) implies that $\sigma_{k+1}$ becomes reducible.

Let us compute the probability $Q$ of the event to find such sequence.
Suppose that generators $\sigma_k$ and $\sigma_{k+1}$ emerge elsewhere in 
the reduced word. We have to push them to the right until they meet the
generator $\sigma_k$ already belonging to the set $I$---see the figure
below

\bigskip

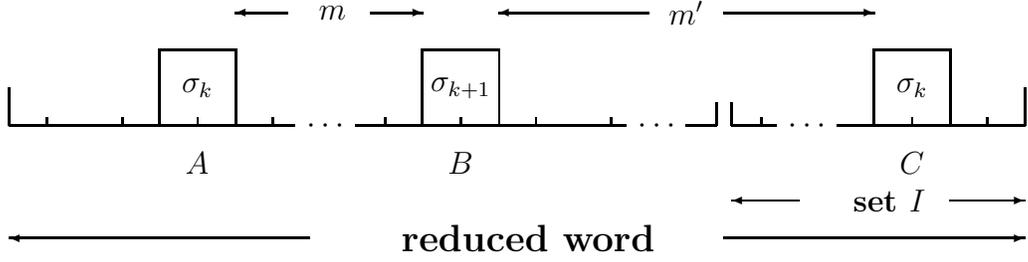
\unitlength=1.00mm
\special{em:linewidth 0.5pt}
\linethickness{0.5pt}
\begin{picture}(150.00,35.00)
\hspace{-1cm}
\put(20.00,20.00){\line(0,1){1.00}}
\put(30.00,20.00){\line(0,1){1.00}}
\put(40.00,20.00){\line(0,1){1.00}}
\put(75.00,20.00){\line(0,1){1.00}}
\put(85.00,20.00){\line(0,1){1.00}}
\put(95.00,20.00){\line(0,1){1.00}}
\put(135.00,20.00){\line(0,1){1.00}}
\put(115.00,20.00){\line(0,1){1.00}}
\put(15.00,25.00){\line(0,-1){5.00}}
\put(70.00,20.00){\framebox(10.00,10.00)[cc]{$\sigma_{k+1}$}}
\put(130.00,20.00){\framebox(10.00,10.00)[cc]{$\sigma_k$}}
\put(150.00,20.00){\line(0,0){0.00}}
\put(150.00,20.00){\line(0,1){5.00}}
\put(35.00,20.00){\framebox(10.00,10.00)[cc]{$\sigma_k$}}
\put(40.00,15.00){\makebox(0,0)[cc]{$A$}}
\put(145.00,20.00){\line(0,1){1.00}}
\put(65.00,20.00){\line(0,1){1.00}}
\put(111.00,23.00){\line(0,-1){3.00}}
\put(109.00,23.00){\line(0,-1){3.00}}
\put(105.00,20.00){\line(1,0){4.00}}
\put(125.00,20.00){\line(1,0){25.00}}
\put(101.00,20.00){\makebox(0,0)[cc]{$\cdots$}}
\put(121.00,20.00){\makebox(0,0)[cc]{$\cdots$}}
\put(75.00,15.00){\makebox(0,0)[cc]{$B$}}
\put(135.00,15.00){\makebox(0,0)[cc]{$C$}}
\put(61.00,20.00){\line(1,0){36.00}}
\put(111.00,20.00){\line(1,0){6.00}}
\put(15.00,20.00){\line(1,0){38.00}}
\put(50.00,20.00){\line(0,1){1.00}}
\put(57.00,20.00){\makebox(0,0)[cc]{$\cdots$}}
\put(100.00,35.00){\vector(-1,0){20.00}}
\put(110.00,35.00){\vector(1,0){20.00}}
\put(63.00,35.00){\vector(1,0){7.00}}
\put(52.00,35.00){\vector(-1,0){7.00}}
\put(58.00,35.00){\makebox(0,0)[cc]{$m$}}
\put(105.00,35.00){\makebox(0,0)[cc]{$m'$}}
\put(132.00,10.00){\makebox(0,0)[cc]{\bf set $I$}}
\put(120.00,10.00){\vector(-1,0){9.00}}
\put(140.00,10.00){\vector(1,0){10.00}}
\put(84.00,5.00){\makebox(0,0)[cc]{\large\bf reduced  word}}
\put(110.00,5.00){\vector(1,0){40.00}}
\put(55.00,5.00){\vector(-1,0){40.00}}
\end{picture}

We proceed in two subsequent steps:
\begin{enumerate}
\item We push the generator $\sigma_k$ located at the point $A$ until it
meets the generator $\sigma_{k+1}$ located at the point $B$. The local
transition probability of such process is $p_1$, where
\be \label{rr:4}
p_1=\frac{2n-6}{2n-1}
\ee
It easy to understand that $p_1$ is the probability to commute a given
generator inside the reduced word with its right neighbour;
\item Completing the first process we push the pair $\sigma_k\sigma_{k+1}$
until it meets the generator $\sigma_k$ located at the point $C$. The
local transition probability of such process is $(p_1\,p_2)$, where
\be \label{rr:5}
p_2=\frac{2n-2}{2n-1}
\ee
$p_2$ is the conditional probability to commute $\sigma_k$ {\it under the
condition} that $\sigma_{k+1}$ commutes as well.
\end{enumerate}

We arrive finally at the equation for $Q(\mu')$:
\be \label{rr:3}
Q(\mu')=\left(\frac{1}{2n}\right)^2\,
\left(\sum_{m,m'=0}^{m+m'\le\mu} p_1^m\,\left(p_1\,p_2\right)^{m'}\right)
\ee

The answer for $Q(\mu')$ in the limit $n\gg 1$ reads:
\be \label{rr:6}
Q(\mu')=\frac{1}{30}-\frac{1}{5}e^{-\frac{5}{2}\mu'}+\frac{1}{6}e^{-3\mu'}
\ee
Moreover, for given $\sigma_k\in I$, not only the sequence $\sigma_k
\sigma_{k+1} \sigma_k$ can be used for braiding relations but also $5$
other sequences (namely $\sigma_k^{-1}\sigma_{k+1}\sigma_k$,
$\sigma_k^{\pm 1} \sigma_{k-1} \sigma_k $, $\sigma_k^{-1} \sigma_{k+1}^{-1}
\sigma_k$, $\sigma_k^{-1}\sigma_{k-1}^{-1}\sigma_k$).

Thus, in Eq.(\ref{dd:8}), $\eta'$ has to be replaced by:
\be \label{rr:7}
\eta''(\mu')=\eta'(\mu')(1+6Q(\mu'))
\ee
while Eq.(\ref{dd:11}) remains unchanged. The results are shown in the lowest
part of Fig.\ref{fig:1} (the points are the numerical simulations for the
 $B_{n+1}$, $n=100$; the full curve corresponds to the
 Eqs.(\ref{rr:7}),(\ref{dd:11})).

At the end of this section let us mention two important facts:
\begin{itemize}
\item For small $N'$ (typically, for $N'<1$ i.e. $N<n$), we get
\be \label{bb:1}
\left(\frac{\left<\mu \right>}{N}\right)_{{\cal LF}_{n+1}}\approx
\left(\frac{\left<\mu \right>}{N}\right)_{B_{n+1}}
\ee
i.e. the "braiding" plays practically no role because the words are too short
to produce sequences such as $\sigma_k\sigma_{k+1}\sigma_k$.
\item In the asymptotic regime $N'\to \infty$ and $\mu'\to\infty$ we get
\be \label{rr1:7}
\eta'=\frac{1}{3},\qquad \eta''=0.4,\qquad \frac{\left<\mu\right>}{N}=0.6
\ee
\end{itemize}

We can now appreciate the impact of the braiding relations. The reductions
are increased by about $20\%$ (from $1/3$ to $0.4$)---see Eq.(\ref{rr1:7})
and, simultaneously, $\left<\mu\right>/N$ is decreased by
about $10\%$. So, in that regime, the groups ${\cal LF}_{n+1}$ and
$B_{n+1}$ do not coincide (eventhough they give the same order of magnitude
for the quantity $\left<\mu\right>/N$). This is consistent with our
conjecture expressed in the paper \cite{den} where we introduced the
concept of the locally free group with "errors" in commutation relations,
${\cal LF}_{n+1}^{\rm err}$. Let us remind that in \cite{den} the coincidence
between the limit behavior of the irreducible words in ${\cal LF}_{n+1}^{\rm
err}$ and $B_{n+1}$ has been reached if we allowed $20\%$ of errors in
commutation relations.

\section{Random Walks with Correlations on Locally Free and Braid Groups}
\label{sec:3}

We come back to the locally free group ${\cal LF}_{n+1}(2)$ and suppose,
now, that the letters are drawn according to the rules described in
the section \ref{subsec:1.1}.

\noindent {\sc The weak correlations (the case "A").}

The effect of correlations "A" amounts to a change of
Eqs.(\ref{dd:7})--(\ref{dd:8}) into:
\be \label{pa:1}
\frac{d\eta'}{dN'}=
2\alpha(1-3\eta')\frac{\beta-\alpha\eta'}{1+\beta-\alpha\eta'}
\ee
and
\be \label{pa:2}
\frac{d\mu'}{dN'}=\beta-\alpha\eta'
\ee
where
\be
\alpha=1-2q_A; \qquad \beta=\sqrt{\frac{1-2q_A}{1+2q_A}}
\ee

Let us explain now how these equations come from.

Using the same line of thought as in the section \ref{subsec:2.1} and
taking into account the normalisation condition (\ref{nor:1}), we get:
\be \label{pa:3}
\frac{d\eta'}{dN'}=(1-3\eta')(1-2q_A)\left(1-\frac{\Pi_2}{\Pi_1}\right)
\ee
and
\be \label{pa:4}
\frac{d\mu'}{dN'}=\Pi_1-\Pi_2
\ee

However, the probabilities, $\Pi_1$ and $\Pi_2$ corresponding to the
"increase" and "decrease" processes ($\Pi_1+\Pi_2=1$) must be computed
again to take into account the correlations. This is done as follows.

1. The case "A" means that we mainly take care of the situation when the
next added letter is the same (with the probability $q_A$) as the previous
added one.

Suppose that, at some  time, we draw the letter $\sigma_k$.
 At the next time step we can add $\sigma_k$ or
$\sigma_k^{-1}$ with the probability\footnote{We draw $\sigma_k$ with
the probability $q_A$ and $\sigma_k^{-1}$ with the probability $q_A$
independent with each others.} $q_A$. Continuing such process we develop a
one--dimensional "random walk"
$$
W_{\rm 1D}^A:\;\left\{\sigma_k\, \sigma_k^{-1}\, \sigma_k\, \sigma_k\,
\sigma_k^{-1}\sigma_k^{-1}\ldots\right\}
$$
with a mean ``lifetime'' $\tau_A=1+2q_A+(2q_A)^2+\ldots =1/(1-2q_A)$.
 This implies the rescaling of the ``time'':
$$
N'\to (1-2q_A)N'
$$
as it can be seen in Eq.(\ref{pa:3}).

2. Another contribution to Eqs.(\ref{pa:3})--(\ref{pa:4}) is connected with
the "mean length", $\left<l\right>$,
of the random chain $W_{\rm 1D}^A$ discussed just above. To
clarify what is $\left<l\right>$ let us consider the following example:

\begin{center}
\begin{tabular}{clcc}
\hline\hline
number of steps & 1D chain & the length  $l$ & probability
\\[5pt] \hline\hline
1 & $\sigma_k$ & 1 & $(1-2q_A)$ \\ \hline
2 & $\sigma_k\sigma_k^{-1}$ & 0 & $q_A(1-2q_A)$ \\
2 & $\sigma_k\sigma_k$ & 2 & $q_A(1-2q_A)$ \\ \hline
3 & $\sigma_k\sigma_k^{-1}\sigma_k=\sigma_k$ & 1 & $q_A^2(1-2q_A)$ \\
3 & $\sigma_k\sigma_k^{-1}\sigma_k^{-1}=\sigma_k^{-1}$ & 1 &
$q_A^2(1-2q_A)$ \\
3 & $\sigma_k\sigma_k\sigma_k=\sigma_k^3$ & 3 & $q_A^2(1-2q_A)$ \\
3 & $\sigma_k\sigma_k\sigma_k^{-1}=\sigma_k$ & 1 & $q_A^2(1-2q_A)$ \\
$\cdots$ & $\cdots$ & $\cdots$ & $\cdots$ \\
\hline\hline
\end{tabular}
\end{center}

The calculation of   $\left<l\right>$ for given $q_A$ and
infinite long random chain $W_{\rm 1D}^A$ leads to the equation
$$
\left<l\right>=\sum_{k=0}^{\infty}a_k\,q_A^k\,(1-2q_A)
$$
where $a_k$ obeys the recursion relations
$$
a_{2k+1}=2\,a_{2k}
$$
$$
a_{2k}-2\,a_{2k-1}=C_{2k}^k
$$
The final answer for $\left<l\right>$ is:
\be \label{pa:5}
\left<l\right>=\sum_{k=0}^{\infty} C_{2k}^k\, q_A^{2k}=
\frac{1}{\sqrt{1-4q_A^2}}
\ee
$\left<l\right>$ is produced during the ``lifetime'' $\tau_A$.
  Thus, Eq.(\ref{dd:8}) should be rewritten as:
\be \label{pa:6}
\frac{1}{1-2q_A}\frac{d\mu'}{dN'}=\left<l\right>-\eta'
\ee
This equation enables us to extract the expressions for $\Pi_1$ (and,
hence, for $\Pi_2=1-\Pi_1$)---see Eq.(\ref{pa:4}) and substitute these
values in Eq.(\ref{pa:3}). Now Eqs.(\ref{pa:1})--(\ref{pa:2}) follow
directly from Eqs.(\ref{pa:3})--(\ref{pa:4}).

Let us stress that naive "time rescaling" in Eq.(\ref{dd:8}), i.e. the
equation
$$
\disp \frac{1}{1-2q_A}\frac{d\mu'}{dN'}=1-\eta'
$$
(i.e., when  $\left<l\right>=1$) leads to a wrong result.

The comparison with numerical simulations is shown in Fig.\ref{fig:2}. For
the group $B_{n+1}$---the lower part of the Fig.\ref{fig:2}---we used the
same recipe as in the section \ref{subsec:2.3} to get the analytic results
(full curve). In the limit $N'\to \infty$ our computations get the answer
(for the group ${\cal LF}_{n+1}$):
\be \label{add:1}
\eta'_{\infty}=\frac{1}{3}\,, \qquad
\left.\frac{\left<\mu\right>}{N}\right|_{\infty}=
\sqrt{\frac{1-2q_A}{1+2q_A}}-\frac{1}{3}(1-2q_A)
\ee
The value $\disp \left.\frac{\left<\mu\right>}{N}\right|_{\infty}$ is a
monotonically decreasing function of $q_A$ when $q_A$ increases from $0$
$\disp \left(\frac{\left<\mu\right>}{N}=\frac{2}{3}\right)$ till its maximal
value $1/2$ $\disp\left(\frac{\left<\mu\right>}{N}=0\right)$. Clearly, the
correlations "A" enhance the reductions.

Let us pay attention to the fact that
\be \label{add:2}
\frac{\left<\mu\right>}{N}\; \mathrel{\mathop{\longrightarrow}
\limits_{N'\to 0}}\; \sqrt{\frac{1-2q_A}{1+2q_A}}<1 \qquad
\mbox {if $q_A\neq 0$}
\ee
where $N'\to 0$ means that $n\gg N\gg 1$. This fact is clearly depicted in
Fig.\ref{fig:3}. One can see that the agreement between Eq.(\ref{add:2})
(full curve) and numerical simulations for the groups ${\cal LF}_{n+1}$ and
$B_{n+1}$ (points) is perfect.

\noindent {\sc The strong correlations (the case "B").}

The correlations "B" are dealing with the situation when the next added
letter to the initial word is $\sigma_{k\pm 1}^{\pm 1}$ (with the
probability $q_B$) if the previous added one is $\sigma_k^{\pm 1}$---see
the section \ref{subsec:1.1} for the definition and the normalisation of
probabilities.

In the spirit of discussion of the case "A" we can describe our process of
successive letters drawing as developing of the 1D Markov chain
$$
W_{\rm 1D}^B:\;\left\{\sigma_k\sigma_{k+1}\sigma_k^{-1}
\sigma_{k-1}^{-1}\sigma_{k-2} \sigma_{k-3}^{-1}\sigma_{k-2}\ldots\right\}
$$
The corresponding "lifetime", $\tau_B$, is $\disp \tau_B=
\frac{1}{(1-4q_B)}$.

The chain $W_{\rm 1D}^B$ can be viewed as a 1D random walk in the "label
space"
$$
k\to (k+1)\to k\to (k-1)\to (k-2)\to (k-3)\to (k-2)\ldots
$$
with an extension around $k$  of order of \
 $\disp 2\sqrt{\tau_B}= \frac{2}{\sqrt{1-4q_B}}$.

Now, if we apply the evolution mechanism of the set $I$, we immediately
realize that that all the generators in a zone of extension
$\disp \frac{2}{\sqrt{1-4q_B}}$ are erased. (In our example: $\sigma_{k+1}$ erases
$\sigma_k$, $\sigma_k^{-1}$ erases $\sigma_{k+1}$, $\sigma_{k-1}^{-1}$
erases $\sigma_k^{-1},\ldots$).

Comparing with the group ${\cal LF}_{n+1}(d)$ and following the remark at
the beginning of section \ref{subsec:2.2}, we can define the new effective
$d=d_{\rm eff}$ by the equation:
\be \label{pb:1}
2\left(d_{\rm eff}-1\right)=\frac{2}{\sqrt{1-4q_B}}
\ee
Moreover, it is easy to see that the probability $\Pi_2$ (reduction
process) is equal to
\be \label{pb:2}
\Pi_2=p_B\,\eta=n\,p_B\,\eta'=\left(\frac{1-4q_B}{2}\right)\eta' \ee We
have used the normalisation (\ref{nor:2}) and supposed that $n\gg 1$.

So, we get:
\be \label{pb:3}
\frac{d\mu'}{dN'}=\Pi_1-\Pi_2=1-2\Pi_2=1-(1-4q_B)\eta'
\ee
For the evolution of $\eta'$, we obtain, after the time rescaling $N'\to
(1-4q_B)N'$, the equation
\be \label{pb:4}
\frac{d\eta'}{dN'}=\left(1-\eta'(2d_{\rm eff}-1)\right)(1-4q_B)
\left(1-\frac{\Pi_2}{\Pi_1}\right)
\ee

 that is easily solved.
 In the limit $N'\to \infty$ we arrive at the following equations
\be \label{pb:5}
\eta'_{\infty}=\frac{1}{2d_{\rm eff}-1}=\frac{1}{\frac{2}{\sqrt{1-4q_B}}+1}
\ee
and
\be \label{pb:6}
\left.\frac{\left<\mu\right>}{N}\right|_{\infty}
=\frac{2+4q_B\sqrt{1-4q_B}}{2+\sqrt{1-4q_B}}
\ee
We can see that $\disp \left.\frac{\left<\mu\right>}{N}\right|_{\infty}$
is a monotonically increasing function of $q_B$ from $2/3$ (for $q_B=0$)
till 1 (for the maximal value $q_B=1/4$).

The correlations in the case "B" increase the length of reduced words in
case of the locally free group ${\cal LF}_{n+1}$. The same behaviour is
seen numerically for the braid group $B_{n+1}$.

In the Fig.\ref{fig:4} we compare the results of numerical simulations for
the group ${\cal LF}_{n+1}(2)$ at $q_B=0.05$ (dots) with our analytic
computations (full line---solutions of Eqs.(\ref{pb:3})--(\ref{pb:4})).

Our numerical computations of the normalised reduced word length, $\mu/N$,
as a function of normalised initial word length, $N/n$, are summarized in
the Fig.\ref{fig:5}. This plot shows the dependence $\mu(N')$ for locally
free and braid groups for both kinds of correlations ("A" and "B"). The
corresponding analytic results are available in all cases except for the
braid group when words are drawn with correlations "B".

\section{A Ballistic Growth Model} \label{sec:4}

We apply the ideas developed above to the investigation of some statistical
properties of a ballistic growth process in 1+1 dimensions.

The standard ballistic deposition can be defined in the following way
\cite{zhang}. Take $n$ columns, of unit width each. A particle, of
unit width and height, is dropped vertically in a randomly choosen column
and sticks upon first contact with the evolving deposit.

Let $h(k,N)$ be the height of the column with the number $k$
($k\in 1,\ldots,n$) after dropping of $N$ particles. The surface of the pile
is determined by the function $h(k,N)$. The change of $h(k,N)$ when one
extra particle is dropped in column $k$ satisfies the following rule:
\be \label{ba:1}
 h(k,N+1)=\max\Bigl\{h(k-1,N),\;h(k,N)+1,\;h(k+1,N)\Bigr\}
\ee
Schematically this rule  corresponds to the following process

\vspace{0.3in}

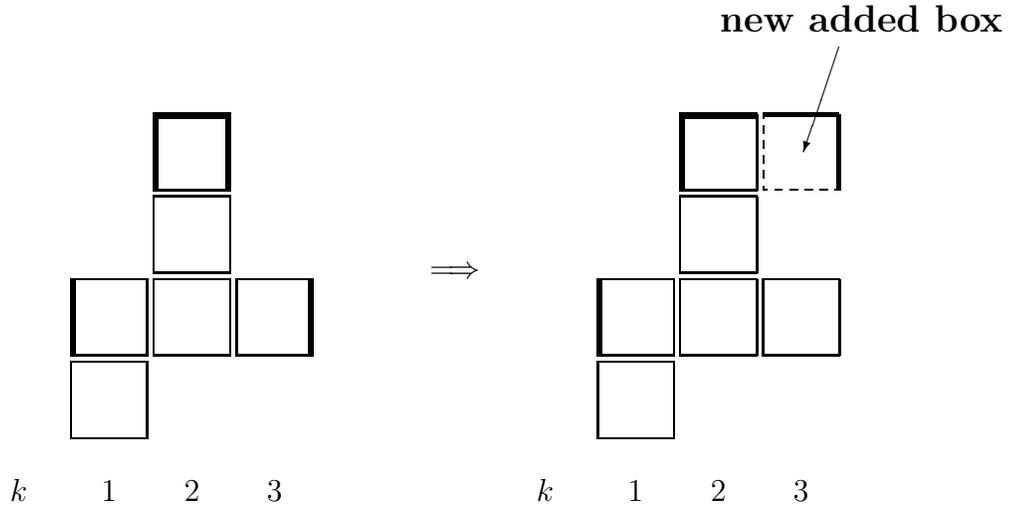
\unitlength=1.00mm
\special{em:linewidth 0.5pt}
\linethickness{0.5pt}
\hspace{1cm}
\begin{picture}(115.00,64.00)
\put(10.00,10.00){\framebox(10.00,10.00)[cc]{}}
\put(10.00,21.00){\framebox(10.00,10.00)[cc]{}}
\put(21.00,21.00){\framebox(10.00,10.00)[cc]{}}
\put(32.00,21.00){\framebox(10.00,10.00)[cc]{}}
\put(21.00,43.00){\framebox(10.00,10.00)[cc]{}}
\put(21.00,32.00){\framebox(10.00,10.00)[cc]{}}
\put(80.00,10.00){\framebox(10.00,10.00)[cc]{}}
\put(80.00,21.00){\framebox(10.00,10.00)[cc]{}}
\put(91.00,21.00){\framebox(10.00,10.00)[cc]{}}
\put(102.00,21.00){\framebox(10.00,10.00)[cc]{}}
\put(91.00,43.00){\framebox(10.00,10.00)[cc]{}}
\put(91.00,32.00){\framebox(10.00,10.00)[cc]{}}
\put(102.00,43.00){\dashbox{1.00}(10.00,10.00)[cc]{}}
\put(61.00,32.00){\makebox(0,0)[cc]{$\Longrightarrow$}}
\put(80.00,21.00){\rule{0.50\unitlength}{10.00\unitlength}} 
\put(91.00,43.00){\rule{0.50\unitlength}{10.00\unitlength}}
\put(91.00,52.50){\rule{10.00\unitlength}{0.50\unitlength}} 
\put(102.00,52.7){\rule{10.00\unitlength}{0.50\unitlength}}
\put(111.70,43.00){\rule{0.50\unitlength}{10.00\unitlength}}
\put(10.00,21.00){\rule{0.50\unitlength}{10.00\unitlength}}
\put(21.00,43.00){\rule{0.50\unitlength}{10.00\unitlength}}
\put(21.00,52.50){\rule{10.00\unitlength}{0.50\unitlength}}
\put(30.50,43.00){\rule{0.50\unitlength}{10.00\unitlength}}
\put(41.50,21.00){\rule{0.50\unitlength}{10.00\unitlength}}
\put(115.00,64.00){\makebox(0,0)[cb]{\large \bf new added box}}
\put(112.00,62.00){\vector(-1,-3){4.67}}
\put(3.00,3.00){\makebox(0,0)[cc]{$k$}}
\put(15.00,3.00){\makebox(0,0)[cc]{1}}
\put(26.00,3.00){\makebox(0,0)[cc]{2}}
\put(37.00,3.00){\makebox(0,0)[cc]{3}}
\put(73.00,3.00){\makebox(0,0)[cc]{$k$}}
\put(85.00,3.00){\makebox(0,0)[cc]{1}}
\put(96.00,3.00){\makebox(0,0)[cc]{2}}
\put(107.00,3.00){\makebox(0,0)[cc]{3}}
\end{picture}

\noindent The "active" box sides (i.e. the sides which can attract the new
particles) are shown in boldface.

Let us slightly modify the rule (\ref{ba:1}) and suppose that:
\be \label{ba:2}
h(k,N+1)=\max\Bigl\{h(k-1,N),\;h(k,N),\;h(k+1,N)\Bigr\}+1
\ee

The prescription (\ref{ba:2}) corresponds to the situation shown below

\vspace{0.3in}

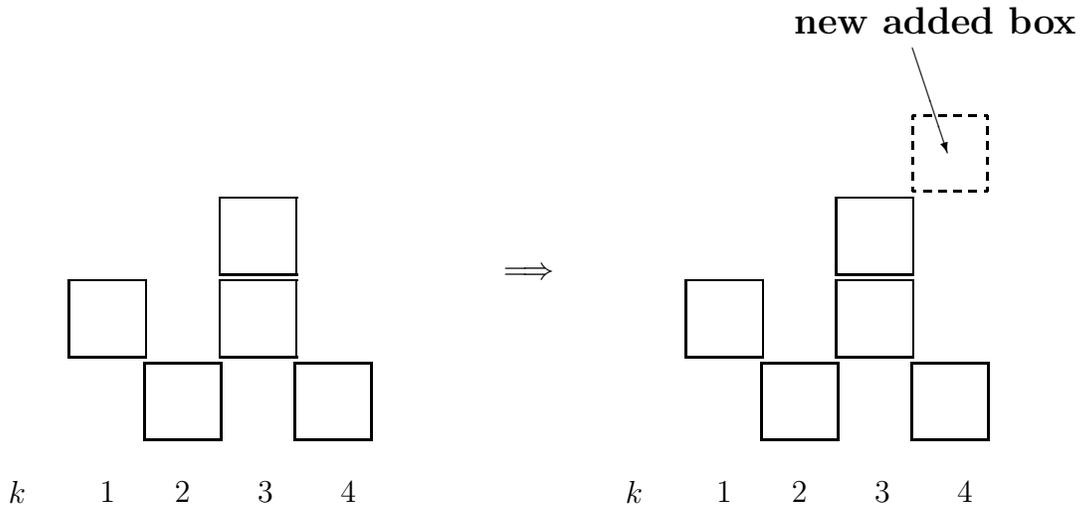
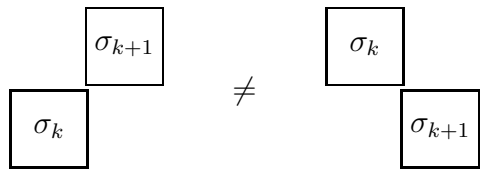
\unitlength=1.00mm
\special{em:linewidth 0.6pt}
\linethickness{0.6pt}
\begin{picture}(132.00,104.00)
\put(10.00,61.00){\framebox(10.00,10.00)[cc]{}}
\put(20.00,50.00){\framebox(10.00,10.00)[cc]{}}
\put(30.00,61.00){\framebox(10.00,10.00)[cc]{}}
\put(40.00,50.00){\framebox(10.00,10.00)[cc]{}}
\put(30.00,72.00){\framebox(10.00,10.00)[cc]{}}
\put(122.00,83.00){\dashbox{1.00}(10.00,10.00)[cc]{}}
\put(71.00,72.00){\makebox(0,0)[cc]{$\Longrightarrow$}}
\put(125.00,104.00){\makebox(0,0)[cb]{\large \bf new added box}}
\put(122.00,102.00){\vector(1,-3){4.67}}
\put(25.00,43.00){\makebox(0,0)[cc]{2}}
\put(36.00,43.00){\makebox(0,0)[cc]{3}}
\put(47.00,43.00){\makebox(0,0)[cc]{4}}
\put(15.00,43.00){\makebox(0,0)[cc]{1}}
\put(3.00,43.00){\makebox(0,0)[cc]{$k$}}
\put(92.00,61.00){\framebox(10.00,10.00)[cc]{}}
\put(102.00,50.00){\framebox(10.00,10.00)[cc]{}}
\put(112.00,61.00){\framebox(10.00,10.00)[cc]{}}
\put(122.00,50.00){\framebox(10.00,10.00)[cc]{}}
\put(112.00,72.00){\framebox(10.00,10.00)[cc]{}}
\put(107.00,43.00){\makebox(0,0)[cc]{2}}
\put(118.00,43.00){\makebox(0,0)[cc]{3}}
\put(129.00,43.00){\makebox(0,0)[cc]{4}}
\put(97.00,43.00){\makebox(0,0)[cc]{1}}
\put(85.00,43.00){\makebox(0,0)[cc]{$k$}}
\put(40.00,10.00){\framebox(10.00,10.00)[cc]{$\sigma_k$}}
\put(50.00,21.00){\framebox(10.00,10.00)[cc]{$\sigma_{k+1}$}}
\put(82.00,21.00){\framebox(10.00,10.00)[cc]{$\sigma_{k}$}}
\put(92.00,10.00){\framebox(10.00,10.00)[cc]{$\sigma_{k+1}$}}
\put(71.00,20.00){\makebox(0,0)[cc]{$\neq$}}
\end{picture}

\noindent It represents the ballistic growth of the pile of unpenetrable
particles still of unit height but of width slightly larger than one: two
particles dropped in neighbouring columns cannot "pass through" each other.

In course of numeric computations we get for the average height of the
pile, $\overline{h(N')}$, the asymptotic value
$$
\frac{\overline{h(N')}}{N'}\approx 4.05
$$
for $N'\equiv N/n \gg 1$; while for the standard ballistic model one
has
$$
\frac{\overline{h(N')}}{N'}\approx 2.13
$$
Thus, the compactness of the pile in our model is about twice
smaller.

The collection of peaks and valleys in our model forms a highly rough
surface, which develops in course of particles dropping. We suppose that
each "time step" corresponds to adding one extra particle to the system.
Recall that $k$ is a peak at some time $N$ if $h(k,N)>\max
\Bigl\{h(k-1,N),h(k+1,N)\Bigr\}$. In what follows we are mainly interested
in computing the average number of peaks, $\eta(N)$. As before, we define
 $\eta'\equiv \eta /n$ and $N'\equiv N/n$.

According to the rule (\ref{ba:2}), two peaks cannot appear in
neighbouring columns and we can easily establish the connection with the
ideas developed above: a particle dropped in the column $k$ can be viewed
as a letter $\sigma_k$ drawn with the probability $1/n$ over the set
$\{\sigma_1,\ldots,\sigma_n\}$ generating the group ${\cal LF}_{n+1}(2)$.
The "hardcore" constraint implies the condition $\sigma_i\sigma_j=
\sigma_j\sigma_i$ if and only if $i\neq j\pm 1$. Note that we deal in this
case with the "semigroup" ${\cal LF}^{+}_{n+1}$ because we do not use the
inverse generators $\sigma_i^{-1}$ and do not consider the reducing
process. From this point of view, our analysis, though analogous, is
simpler than for the whole group ${\cal LF}_{n+1}$.

Thus, we can easily deduce that {\it the set of peaks is analogous to the
set of reducible letters $I$} and is reminiscent of the enumeration of
"partially commutative monoids" known in combinatorics \cite{monoids}.

Suppose that two neighbouring peaks are separated by the horizontal
interval of length $i\ge 1$ and $n_i$ is the number of such intervals. Now
we are in position to write the recursion relations for the process $N'\to
N'+1/n$:
\be \label{ba:3}
\left\{\begin{array}{ll}
\eta'\to \eta' & \mbox{occurs with probability $\disp \; \Pi'_0=\frac{1}{n}
\left(\eta+2\sum_{i\ge 2}n_i\right)$} \medskip \\
\disp \eta'\to\eta'-\frac{1}{n} & \mbox{occurs with probability
$\disp\; \Pi'_{-}=\frac{n_1}{n}$} \medskip \\
\disp \eta'\to\eta'+\frac{1}{n} & \mbox{occurs with probability $\disp \;
\Pi'_{+}= \frac{1}{n}\left(\sum_{i\ge 3}n_i\,(i-2)\right)$}
\end{array} \right.
\ee
where the conservation condition implies that
$$
\Pi'_0+\Pi'_{-}+\Pi'_{+}=1
$$
The sum rules (\ref{dd:1})--(\ref{dd:2}) remain to be unchanged. Comparing
Eqs.(\ref{ba:3}) with Eqs.(\ref{dd:30}) we find that only $\Pi'_0$ differs
from $\Pi_0$.

In terms of $\eta'$ and $N'$, we get the simple ordinary differential
equation for the mean value
\be \label{ba:6}
\frac{d\eta'}{dN'}=1-3\eta'
\ee
 The solution of Eq.(\ref{ba:6}) reads
\be \label{ba:7}
\eta'= \frac{1}{3}\left(1-e^{-3N'}\right)
\ee
So, asymptotically ($N'\to\infty$), we get that $1/3$ of the columns are
peaks.

Let us extend our consideration to the case of unpenetrable particles of
widths slightly larger than $(d-1)\times$(unit particle width). It means
that now the "hardcore" condition forces us to consider the generators with
the following commutation relations: $\sigma_i\sigma_j=\sigma_j\sigma_i$ if
and only if $|i-j|\ge d$ (where $d\ge 2$). This situation is shown
schematically below:

\bigskip

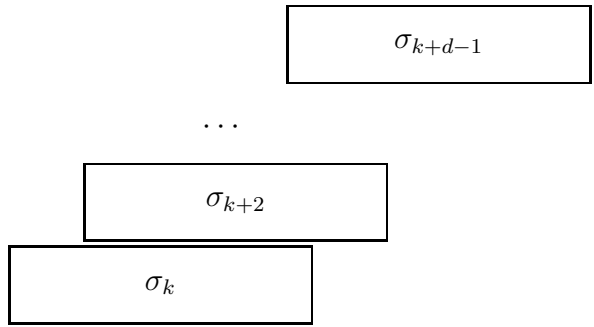
\unitlength=1.00mm
\special{em:linewidth 0.6pt}
\linethickness{0.6pt}
\begin{picture}(110.00,44.00)
\put(20.00,2.00){\framebox(40.00,10.00)[cc]{$\sigma_k$}}
\put(30.00,13.00){\framebox(40.00,10.00)[cc]{$\sigma_{k+2}$}}
\put(48.00,28.00){\makebox(0,0)[cc]{$\cdots$}}
\put(57.00,34.00){\framebox(40.00,10.00)[cc]{$\sigma_{k+d-1}$}}
\end{picture}

It is a simple matter to see how Eqs.(\ref{ba:3})--(\ref{ba:7}) are
changed. We quote the final result only:
\be \label{ba:8}
\eta'=\frac{1}{2d-1}\left(1-e^{-(2d-1)N'}\right)
\ee
The case $d=2$ corresponds to Eq.(\ref{ba:7}).

As it has been done for the locally free group ${\cal LF}_{n+1}(2)$, we are
looking at the changes in $\eta'$ (Eq.(\ref{ba:7})) when we allow the
correlations between the subsequently dropped particles (see the section
\ref{sec:3} for details).

\noindent {\sc The weak correlations (the case "A").}

If we draw $\sigma_k$ at some moment of time, $N$, then at the next moment
of time, $(N+1)$, we have the following situation:
$$
\left\{\begin{array}{l}
\mbox{The generator $\sigma_k$ appears with the probability $q_A$} \\
\mbox{Any generator $\sigma_l$ ($l\ne k$) appears with the probability
$p_A$} \end{array} \right.
$$
Due to the absence of inverse generators, the normalisation reads now:
\be \label{bapa:1}
q_A+(n-1)\,p_A=1
\ee
(compare to Eq.(\ref{nor:1})).

The recursion rules for the process $N'\to N'+1/n$ are now:
\be
\left\{\begin{array}{ll}
\Delta\eta'=0   & \mbox{occurs with probability $\disp \Pi'_0=q_A+
(\eta-1)p_A+ 2\sum_{i\ge 2}n_i\,p_A$} \medskip \\
\disp \Delta\eta'= -\frac{1}{n} & \mbox{occurs with probability $\Pi'_{-}=
n_1\,p_A$} \medskip \\
\disp \Delta\eta'= +\frac{1}{n} & \mbox{occurs with probability $\disp
\Pi'_{+}=\sum_{i\ge 2}n_i\,(i-2)\,p_A$}
\end{array}\right.
\ee
So, we get
\be \label{bapa:2}
\frac{d\eta'}{dN'}=(1-q_A)(1-3\eta')
\ee
and
\be \label{bapa:3}
\eta'=\frac{1}{3}\left(1-e^{-3(1-q_A)N'}\right)
\ee

Asymptotically, again $1/3$ of the columns are peaks and the effect of
correlations leads here only to the time rescaling $N'\to (1-q_A)N'$.

\noindent {\sc The strong correlations (the case "B").}

If we draw $\sigma_k$ at some moment of time, $N$, then at the next moment
of time, $(N+1)$, we have the situation:
$$
\left\{\begin{array}{l}
\mbox{The generator $\sigma_{k\pm 1}$ appears with the probability $q_B$}
\\
\mbox{Any generator $\sigma_l$ ($l\ne k\pm 1$) appears with the probability
$p_B$}
\end{array} \right.
$$
The normalisation is:
\be \label{bapb:1}
2q_B+(n-2)\,p_B=1
\ee

As it has been shown already in section \ref{sec:3}, the effect of such
correlations leads to replacement of $d=2$ by some effective $d_{\rm eff}$.
According to Eq.(\ref{bapb:1}), we get:
\be \label{bapb:2}
d_{\rm eff}= \frac{1}{\sqrt{1-2q_B}}+1
\ee
In addition, the time has to be rescaled as $N'\to (1-2q_B)\,N'$.
Finally we arrive at the following linear differential equation
\be \label{bapb:3}
\frac{d\eta'}{dN'}=(1-2q_B)\left(1-(2d_{\rm eff}-1)\eta'\right)
\ee
which has the solution
\be \label{bapb:4}
\eta'=\frac{1}{2d_{\rm eff}-1}\left(1-e^{-(2d_{\rm eff}-1)(1-2q_B)N'}
\right)
\ee

It is worthwile to notice that, in addition to the time rescaling, the
 correlations ``B'' also lead to a change of the asymptotic value of $\eta'$.

Comparison of Eqs.(\ref{bapa:3})--(\ref{bapb:4}) with numerical simulations
($n=1000$) shows quite good agreement---see Fig.\ref{fig:6} for
correlations of type "A": (a) $q_A=0.2$, (b) $q_A=0.5$ and Fig.\ref{fig:7}
for correlations of type "B": (a) $q_B=0.1$, (b) $q_B=0.2$.

\section{Conclusion}

The investigation of statistical properties of random walks on braid and
locally free groups is undertaken due to the following reasons:

1. On the basis of performed investigation we are going to construct the
simple mean--field Flory--type theory of interacting braided random walks
(bunches of "directed polymers") with nonabelian topology in 1+1
dimensions.

2. The minimal irreducible length of the braid (i.e. the reduced word)
can be served as a well defined characteristic of the "complexity" of
knots constructed on the basis of braids. Thus, our study could be regarded
as a basis for investigation of the limit behavior of knot and link
topological invariants when the length of the corresponding braid tends
to infinity, i.e., when the braid "grows".

3. We believe that the application of the locally free group in the theory
of ballistic aggregation could be used: (i) in the consideration of
statistical and relaxational properties of "sandpile models" exhibiting
SOC--behavior\footnote{SOC is the abbrevation of "self--organised
criticality".}; (ii) in the microscopic descriprion of the surface growth
phenomena.

Of course, our investigation is far from being complete. For instance, the
scaling--like considerations of correlations "A" in locally free and braid
group should be justified from the point of view of standard
renormalisation group technique; special care should be taken for analytic
consideration of correlations "B" in braid group and so on.

However we would like to finish the paper on the optimistic note. Let us
express the hope that the problem of discovering the integrable models
associated with the proposed locally free groups and developing the
corresponding conformal field theory could help to establish the bridge
between statistics of random walks on the noncommutative groups, spectral
theory on multiconnected Riemann surfaces, topological field theory and
statistics of rough surfaces in models of ballistic aggregation and
sandpile growth.

\newpage

\section*{Figure Captions}

\begin{fig}
The normalised length, $<\mu>/N$, of the reduced word as a function of 
 the length  of the initial word, $N/n$, for locally free and braid groups. 
 Words are drawn without any correlations.
\label{fig:1}
\end{fig}

\begin{fig} The same as Fig.\ref{fig:1} except that  words are drawn
with correlations "A" ($q_A=0.1$).
\label{fig:2}
\end{fig}

\begin{fig}
 The limit of  $<\mu>/N$  when  $0<N/n\ll 1$ is plotted  
 as a function of the probability   $q_A$, for locally free and braid groups.
  Words are drawn with correlations "A".
\label{fig:3}
\end{fig}

\begin{fig} The same as Fig.\ref{fig:1} except that words are drawn
with correlations "B" ($q_B=0.05$). (Analytic computations for the 
group $B_n$ are absent). 

\label{fig:4}
\end{fig}

\begin{fig}
The plot shows the dependence $\mu(N')$ for locally free and braid groups
for both kinds of correlations ("A" and "B"). The corresponding analytic
results are available in all cases except for the braid group when words
are drawn with correlations "B".
\label{fig:5}
\end{fig}

\begin{fig}
Dependence of normalised amount of the surface peaks, $\eta'$, on the
normalised number of "pile volume", $N'$, in the ballistic aggregation
model. Particles are dropped with correlations "A".
\label{fig:6}
\end{fig}

\begin{fig}
Dependence of normalised amount of the surface peaks, $\eta'$, on the
normalised number of "pile volume", $N'$, in the ballistic aggregation
model. Particles are dropped with correlations "B".
\label{fig:7}
\end{fig}

\end{document}